\documentclass[sigconf]{acmart}

\AtBeginDocument{%
  \providecommand\BibTeX{{%
    \normalfont B\kern-0.5em{\scshape i\kern-0.25em b}\kern-0.8em\TeX}}}

\copyrightyear{2023}
\acmYear{2023}
\setcopyright{cc}
\setcctype[4.0]{by}
\acmConference[AM '23]{Audio Mostly 2023}{August 30-September 1,
2023}{Edinburgh, United Kingdom}
\acmBooktitle{Audio Mostly 2023 (AM '23), August 30-September 1, 2023,
Edinburgh, United Kingdom}\acmDOI{10.1145/3616195.3616196}
\acmISBN{979-8-4007-0818-3/23/08}

\acmConference[AM ’23]{AudioMostly 2023}{August 30--September 01,
  2023}{Edinburgh, UK}
\acmPrice{15.00}
\acmISBN{979-8-4007-0818-3/18/06}

\begin{document}

\title{FM Tone Transfer with Envelope Learning}

\author{Franco Caspe}
\email{f.s.caspe@qmul.ac.uk}
\affiliation{%
  \institution{Centre for Digital Music, \\Queen Mary University of London}
  \country{United Kingdom}
}

\author{Andrew McPherson}
\email{andrew.mcpherson@imperial.ac.uk}
\affiliation{%
 \institution{Dyson School of Design Engineering, Imperial College London}
 \country{United Kingdom}}

\author{Mark Sandler}
\email{mark.sandler@qmul.ac.uk}
\affiliation{%
  \institution{Centre for Digital Music,\\Queen Mary University of London}
  \country{United Kingdom}
}

\renewcommand{\shortauthors}{Caspe et al.}

\begin{abstract}
Tone Transfer is a novel deep-learning technique for interfacing a sound source with a synthesizer, transforming the timbre of audio excerpts while keeping their musical form content. Due to its good audio quality results and continuous controllability, it has been recently applied in several audio processing tools. Nevertheless, it still presents several shortcomings related to poor sound diversity, and limited transient and dynamic rendering, which we believe hinder its possibilities of articulation and phrasing in a real-time performance context.

In this work, we present a discussion on current Tone Transfer architectures for the task of controlling synthetic audio with musical instruments and discuss their challenges in allowing expressive performances.
Next, we introduce Envelope Learning, a novel method for 
designing Tone Transfer architectures that map musical events using a training objective at the synthesis parameter level. Our technique can render note beginnings and endings accurately and for a variety of sounds; these are essential steps for improving musical articulation, phrasing, and sound diversity with Tone Transfer.
Finally, we implement a VST plugin for real-time live use and discuss possibilities for improvement.
\end{abstract}

\ccsdesc[500]{Applied computing~Sound and music computing}
\ccsdesc[500]{Computing methodologies~Neural networks}
\ccsdesc[500]{Human-centered computing~Interaction techniques}
\keywords{Neural Networks, Synthesis Control, Musical Instrument Interaction, Mapping}



\maketitle

\section{Introduction}

Synthesizers can be very expressive instruments, whether controlled by the ubiquitous keyboard \cite{pinch_SOCIAL_1998}, by augmented instruments, or instrument-like interfaces  \cite{bevilacqua_Augmented_2012,moro_Performer_2020,mcpherson_TouchKeys_2012}, or by whole new sets of gestures enabled by novel controllers \cite{tahiroglu_AIterity_2021,dahlstedt_Physical_2017}.

Recent developments in integrating Deep Neural Networks (DNNs) with audio generators have renewed interest in using the unaltered audio of a musical instrument as a control source for a synthesizer.
One such example is the DDSP architecture and its derivatives \cite{engel_DDSP_2020,hayes_Neural_2021,caspe_DDX7_2022, shan_Differentiable_2022}, that allows for real-time control of a synthesizer using a set of features extracted from an input audio signal. It has been used to develop various creative timbre transformation applications, which we collectively refer to as Tone Transfer applications. \cite{carney_Tone_2021,qosmo_Neutone_2023,bytedance_Mawf_2023,googlemagenta_DDSPVST_2023}.

We situate the scope of our work on audio-based synthesis control for real-time performances, looking at sonic diversity and synthesizer phrasing and articulation. These essential components of musical expression have been thoroughly studied for composition with MIDI for decades \cite{wessel_Control_1987,bresin_Articulation_2000,wu_MIDIDDSP_2022} but we argue that they open new challenges and possibilities when considering an audio-based control approach. Transients at the beginnings of notes and the transitions between notes play a vital role in defining the continuity and flow of musical phrasing. 
We argue that a continuous control approach such as Tone Transfer could potentially learn mappings that capture beginnings, endings, and the links between notes during performance, generating musically articulated synthetic sounds.

In this work, we begin by examining the challenges faced by existing Tone Transfer architectures when it comes to effectively supporting aspects of musical expression such as phrasing, articulation, and sonic diversity. We argue that these challenges are primarily linked to the training methods employed and the commonly used synthesis models. 
Next, we propose Envelope Learning as a method to circumvent these issues. This technique revolves around designing Tone Transfer architectures that focus on matching synthesis parameters instead of audio features. Since our models learn musical events at the level of synthesis control, they can reproduce quick changes in sound, such as the start and end of musical notes, which are essential for musical phrasing.

We train the models to learn different tones by using patches from a well-known FM synthesizer, which provides a diverse range of sounds to work with.
Finally, we implement our models on an audio plugin for real-time performances and reflect on its performance and possibilities for improvement.
For training and deploying source code, see the online supplement \footnote{\url{https://fcaspe.github.io/fmtransfer}}. We expect our models to complement existing Tone Transfer architectures and offer further performance possibilities for live use and sound design.

\section{Background}

\subsection{Synthesis control with audio signals}

Audio-based control in synthesis has a longstanding history, exemplified by pioneering instruments like the Roland guitar \cite{lahdeoja_Approach_2008}. In recent times, a prevalent technique involves using onset detectors and fundamental frequency trackers, enabling control of a synthesizer through MIDI signals \cite{derrien_very_2014}. This method facilitates the translation of audio input into synthesized sounds, offering a versatile approach to musical control.

However, generating MIDI triggers through explicit note onset detection introduces a bottleneck in the gestural channel \cite{jack_rich_2017} between the instrument and the synthesizer which may hinder an expressive performance. In that regard, timbre characteristics related to musical phrasings, such as variations in dynamics, frequency spectrum, amplitude envelope, and attack transients \cite{olsen_What_2016}, are compressed into a single scalar velocity value. In MIDI, phrasing is implicitly represented through the timing and velocities of a sequence of note events. Audio-to-MIDI converters typically identify single notes at a time without consideration of longer sequences, so any temporal jitter or inaccuracy in dynamics could result in a disjointed sense of phrasing when the MIDI sequence is replayed by a synthesizer.

In addition to MIDI-based control, alternative strategies have been investigated, involving the extraction of continuous features from the audio signal of a musical instrument to control synthesis processes. In prior work, audio signals from instruments have been utilized as oscillators \cite{popel_Audio_2005}. This technique employs the audio signal itself as an oscillator for generating synthesized sounds, a special case of an Adaptive Digital Audio Effect \cite{lazzarini_Adaptive_2007,verfaille_Adaptive_2006}. However, this approach restricts the method's versatility as it binds the sonic characteristics of the input directly to the output limiting its range of sonic possibilities.

Continuous control offers the possibility of better supporting musical expression. Interestingly, by closely analyzing how notes from an audio source are intertwined, we may also be able to facilitate longer phrasing arcs on the synthesizer.
The problem now resides in navigating the complexity of the mapping design. What are the features we should extract, and how should we associate them to support a variety of synthetic sounds? There are many degrees of freedom, and the strategy becomes much less evident \cite{rasamimanana_Conceptual_2012}.

One possible answer can be found in the work of Levitin et al., where they proposed a valuable framework for analyzing the processes involved in a musical event control \cite{levitin_Control_2002}.
They outline distinct stages of control within a musical event including the beginning, middle, ending, and terminus of the event, and highlight that Digital Musical Instruments (DMIs) often provide greater control over the middle.

In this context, different beginnings and endings can encompass musical articulation and serve as vital contextual links within musical phrases \cite{lindemann_Music_2007}.
The audio signal contains this important information within a very short duration and may require direct attention and specific handling to accurately capture and preserve these critical elements. These insights underscore the need for a focused approach to address beginnings and endings explicitly.

\subsection{Differentiable Signal Processing}

The seminal work of Van Den Oord et. al. \cite{oord_WaveNet_2016}, spawned a novel approach for data-driven audio generation and control called Neural Audio Synthesis. In this context, Deep Neural Networks (DNN) learn complex synthesizers from audio corpora, that can be used for composition \cite{engel_Neural_2017}, singing voice control \cite{wang_Neural_2019} timbre transformation \cite{huang_TimbreTron_2019} and synthesizer parameter estimation \cite{chen_Sound2Synth_2022}, to name a few applications.

Engel et. al. \cite{engel_DDSP_2020} proposed a method called Differentiable Signal Processing (DDSP) that combines neural networks and DSP modules, such as synthesizers and audio effects, allowing an error signal to be backpropagated through them.
This approach enables joint training of the whole pipeline, effectively biasing the network to learn to control the DSP modules. It allows efficient sound generation with DNN models that can comfortably run in real-time on a CPU \cite{ganis_Realtime_2021} and yield impressive results on a variety of differentiable synthesis architectures for musical instrument \cite{hayes_Neural_2021,caspe_DDX7_2022,shan_Differentiable_2022} and singing voice rendering \cite{wang_Neural_2019,wu1_DDSPbased_2022}.

\subsection{Tone Transfer}

Tone Transfer \cite{carney_Tone_2021} is a promising application enabled by DDSP for audio-based control of synthesizers.
The supporting architecture, called DDSP Decoder \cite{engel_DDSP_2020}, learns to control parameters of a synthesizer, conditioned by a frame-wise fundamental frequency ($F_0$) and loudness sequences extracted from an input audio signal. These characteristics are instrument-agnostic and relate uniquely to musical form; during inference, the model can support any musical instrument signal that contains a tractable $F_0$.

A continuously controllable synthesizer such as the DDSP Decoder can potentially deal with the fine-grained characteristics of note beginnings and endings, essential for phrasing. Nevertheless, we note that certain design decisions related to its architecture and training methods may hinder phrasing, articulation, and sound diversity in a performance setting.

One problem is related to the training process, which aims to resynthesize an audio corpus of a particular instrument from a set of $F_0$ and loudness conditioning sequences, guided by the Multiscale Spectrogram Loss \cite{wang_Neural_2019,turian_Sorry_2020}. This involves a trade-off between time and frequency resolution \cite{ricaud_optimally_2014}, affecting the model's capacity of discerning and synthesizing accurate instrument onsets, which typically happen in the order of tens of milliseconds \cite{vos_perceptual_1981} and are essential to convey distinct articulation and build musical phrases. 

Transient rendering is also affected by the synthesizer architectures typically employed in DDSP decoders. In the majority of the cases, a harmonic source such as a harmonic synth \cite{engel_DDSP_2020}, a waveshaper \cite{hayes_Neural_2021} or a wavetable \cite{shan_Differentiable_2022} is paired with a noise synthesizer in a setting that resembles a Spectral Modelling Synthesizer \cite{serra_Spectral_1990}. This configuration is usually not sufficient for an accurate representation of transients \cite{daudet_Transients_2001,verma_Extending_2000}.

Regarding sound diversity, we note that the resynthesis objective implicitly ensures a high correlation between the input and output loudness, as indicated in the original paper \cite{engel_DDSP_2020}. 
Since different musical instruments have different loudness profiles, in many cases performers expect the dynamic characteristics of the generated audio to be different from those of the input. Losing this degree of freedom may make the learned timbre track the dynamics of the input too closely, producing unnatural sounds and limiting sonic diversity.

Another issue is related to the availability of training data. Single musical instrument datasets are difficult to collect \cite{li_urmp_2019}, and in many cases, $F_0$ may not be easy to extract, especially for synthetic sounds. This also limits the amount and type of sounds that can be synthesized with Tone Transfer.

Finally, it is worth noting that sound design practitioners are not familiar with the spectral modeling synthesizers typically employed for Tone Transfer. A well-known architecture with interpretable parameters allows performers to intervene in the synthesis process and manipulate results enhancing the possibilities of pre-trained models \cite{caspe_DDX7_2022}.

\subsection{FM Synthesis}

Frequency Modulation (FM) synthesis is a well-known method to generate complex sounds from a compact set of synthesis parameters \cite{chowning_Synthesis_1973}. One of the best-known implementations is the Yamaha DX7, which utilizes a well-established linear FM synthesis architecture, that has been used in other works for applications such as sound matching and neural audio synthesis \cite{chen_Sound2Synth_2022,caspe_DDX7_2022}.

The DX7 generates its distinctive sound using six frequency-modulated sinusoidal oscillators. Programming the synthesizer involves configuring a patch that specifies various parameters for each oscillator. These parameters include the routing, which determines how the oscillators are interconnected (e.g., in a stacked or additive manner), the frequency ratios of the oscillators relative to the played note, as well as the Attack-Decay-Sustain-Release (ADSR) parameters of its Envelope Generators (EGs). 

During audio rendering, the oscillator's frequency ratios and routing remain fixed. Instead, the sound dynamics are primarily controlled by the ADSR envelopes. These envelopes modulate the output levels of each oscillator, influencing either their volume or modulation index, depending on their interconnection. 
Sound design on the DX7 involves configuring the routing, frequency ratios, and ADSR parameters of the EGs.

\section{Method}

Existing Tone Transfer architectures have shown the ability to learn relationships between control inputs and synthesizer features. However, we have observed certain limitations in terms of transient generation and sonic diversity that could restrict the performative possibilities of the models.

We propose an alternative design method for a model that learns relationships from a dataset of synthesis control signals extracted from synthesizer patches and designed following the musical event control model described by Levitin et al. \cite{levitin_Control_2002}. The model learns to render note beginnings, middles, and endings directly from a continuous control source.
We use an FM synthesizer based on the Yamaha DX7 for which there is a wide variety of sounds available on the web \cite{turian_One_2021}.

\begin{figure*}[ht]
\includegraphics[width=\textwidth]{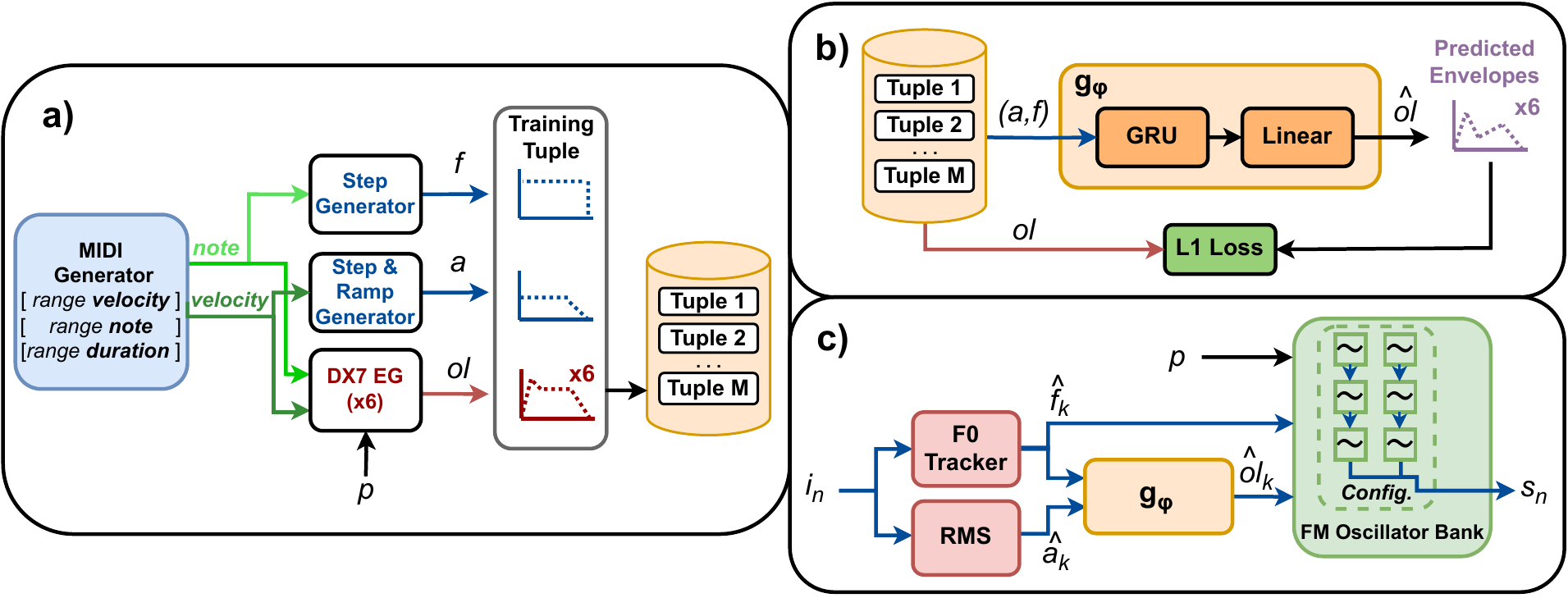}
\Description[Three diagrams depicting each one of the design steps.]{The first diagram shows a generator rendering envelopes. The second one shows the training process. The third one shows a use case within a Tone Transfer pipeline.}
\centering
\caption{Design steps for our Tone Transfer system. a) we create a synthetic dataset of aligned sequences $(a,f,ol)$. $a$ and $f$ model the frame-wise amplitude and $F_0$ trajectories of a monophonic audio signal, while $ol$ are the oscillator output levels of an FM synthesizer programmed with a patch $p$. b) We train a Recurrent Neural Network model $g_{\phi}$ to learn the correspondences between the features $a,f$ and the controls $ol$ reflected in the dataset. c) We deploy the RNN into a Tone Transfer pipeline. In this context, $g_{\phi}$ processes frame-wise input features from real audio $i_n$, and controls the envelopes of an FM oscillator bank configured according to $p$.}
\label{fig:method}
\end{figure*}

We divide our approach into three stages, shown in Figure \ref{fig:method}, namely a dataset generation step that creates event-aligned sequences from synthesizer patches, a training step we call Envelope Learning that learns a mapping function $g_{\phi}$ between these sequences, and an inference step where we deploy trained models into a Tone Transfer pipeline and use them to control an FM oscillator block with audio signals.

Our current research shares similarities with our previous work, where we utilize a neural network to control oscillator amplitudes of an FM synthesizer based on a sequence of audio signal features \cite{caspe_DDX7_2022}. However, in contrast to our earlier approach, we introduce a new design strategy that (1) avoids the reconstruction objective and MSS loss, allowing decoupled dynamics (2) learns an input-to-output mapping at the level of short frames of signal, allowing for accurate transients, and avoiding the use of differentiable synthesis components, and (3) does not require an audio corpus for training, and instead can learn from a patch collection of the FM synthesizer.

\subsection{Dataset Generation}\label{section:dataset}

In this step, we create a dataset of $M$ training tuples $(a^i,f^i,ol^i), i=1,...M$, with $a = a_1,...a_K \in \mathbb{R}$ and $f = f_1,...f_K \in \mathbb{R}$ being sequences of length $K$ modeling amplitude and fundamental frequency of a monophonic audio input respectively. $ol = ol_1,...ol_K \in \mathbb{R}^6$ represent the linear output level envelopes of the six FM oscillators, that we extract from a synthesizer patch. For training, we use $(a,f)$ as input sequences to our model, and $ol$ as supervision.

In order to generate the input sequences $(a^i,f^i)$ we take into consideration the model proposed by Levitin et al., \cite{levitin_Control_2002}. To simplify the dataset generation process, we only consider separate notes as musical events.
For our Tone Transfer use case, an explicit note beginning is determined by a sudden change in the input amplitude contour $a$ and a valid $F_0$ detected in $f$. During the middle, the amplitude and fundamental frequency are sustained over time. Finally, the ending of a note is characterized by a decay trajectory in amplitude, while the $F_0$ remains valid until the terminus.
Considering this, we can model our amplitudes $a$ with a trapezoid generator, that is, a step generator plus a decay ramp. The fundamental frequency contour of a note can be represented with a step generator.

To obtain $ol^i$, we use a Python implementation of the Yamaha DX7 ADSR Envelope Generators (EGs) adapted from a well-known emulator \cite{gauthier_Dexed_2023}. These EGs can be programmed with a synthesizer patch $p$ and actuated through MIDI to obtain the amplitude envelope sequences of the six oscillators.

The dataset generation starts with a designer selecting a synthesizer patch $p$ they want to enable for Tone Transfer. We program the ADSR parameters of the EGs with $p$ and generate a set of MIDI notes of random duration and with random velocity and note values.

For each note, we obtain the oscillator envelope sequences $ol^i$, and create the aligned input sequences $(a^i,f^i)$ following a simple set of rules.
When a "NOTE ON" message is received, we generate a step response with an amplitude proportional to the velocity and note value for $a$ and $f$ respectively. After the "NOTE OFF" event is received, a linear decay ramp is rendered in $a$ until the last oscillator envelope in $ol$ reaches zero. During this time $f$ remains valid and then is set to zero. 

Next, the input sequences are normalized between $[0,1]$, establishing a linear range of $a$ and $f$ corresponding to MIDI velocity and MIDI notes values respectively.
The oscillator envelope sequences fall in the range of $[0,2]$; they are also normalized to a range within $[0,1]$.
Finally, all sequences are padded with zeroes before and after so that each training tuple features the same length.

\begin{figure}[ht]
\centering
\includegraphics[width=0.5\textwidth]{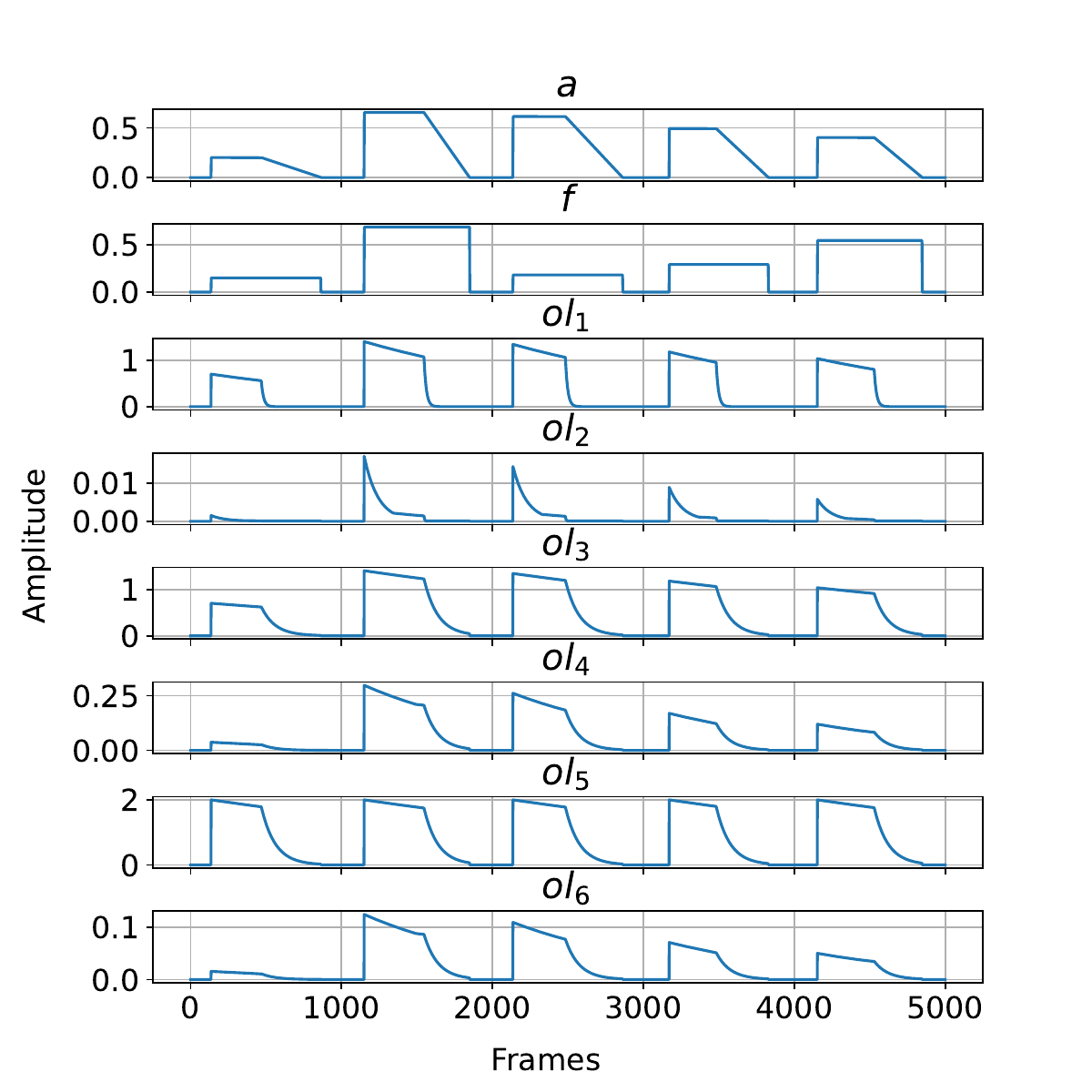}
\Description[Stacked plots of the two inputs and the six oscillator envelope outputs.]{The stacked plot shows the continuous amplitude and f0 signals in sync with the six oscillator envelopes in the dataset.}
\caption{Five training tuples of a dataset extracted from the "E. PIANO 1" patch. Showing its corresponding input sequences $a$ and $f$, and the synchronized envelopes $ol$.}\label{fig:dataset}
\end{figure}

The end result is a dataset that aligns two characteristics of musical events ($a$ and $f$) to synthesizer controls ($ol$) that can render a specific timbre obtained from the patch.
Since the input $a$ is proportional to the velocity, note beginnings are characterized by different amplitude discontinuities which in turn, are aligned with the oscillator envelopes $ol$ that render different onsets. Middles are mostly aligned with the decay and sustain parts of $ol$, and the decaying sections of note endings are synchronized with the release sections of the envelopes. Figure \ref{fig:dataset} shows a plot of the first six training instances of a dataset generated from the "E. PIANO 1" patch, a well-known DX7 electric piano patch, illustrating the synchrony between inputs and oscillator envelopes.

It is important to recognize that musical instrument notes can often display ambiguous behavior, and it is not always the case that a decrease in amplitude indicates the end of a note or a sustained amplitude indicates the middle. In a causal setting for real-time use, we cannot be sure that a note is ending even if there is a decay amplitude trajectory in the input.
Although the input trajectories used in this setting may not fully represent a real-world scenario, they are valuable in demonstrating the proof of concept and analyzing opportunities for improvement.

\subsection{Envelope Learning}

To implement our system, we need a neural network model that can learn the temporal relationships between the inputs $(a,f)$ and the oscillator envelopes $ol$, rendering the attack and decay sections of the control sequences after a discontinuity in the input is detected, and generating note ends accordingly when a decay trajectory is detected in the input.

To this end, and following the design of other Tone Transfer architectures, we employ a model that features a stateful Gated Recurrent Unit (GRU) and a linear layer as output layer.
The GRU is a causal model that works frame-by-frame, is conditioned by the $a$ and $f$ sequences, and learns the relationships between current and past inputs, producing a hidden state that is projected with the linear layer into six controls for the oscillators. We denote the neural net as the parameterized function $g_{\phi}$ as shown in Eqn. \ref{eq:rnn}, where $k$ denotes the frame index. 
\begin{equation}\label{eq:rnn}
    \hat{ol}_k = g_{\phi}(a_k,f_k)
\end{equation}

Since we do not employ audio during our training process, we train the network by conditioning it with $a$ and $f$, and using the oscillator envelopes $ol$ from the dataset as supervision. We use the L1 Loss between the oscillator envelope predictions and ground truth as the minimization objective: $L = ||ol - \hat{ol}||_1$. We call this process Envelope Learning.

The L1 loss aims to match every single frame that is generated by the network directly with the ground truth. This is unlike the DDSP-based methods that learn to control envelopes \textit{indirectly} by employing a resynthesis objective, a spectrogram audio loss, and noise synthesizers. This results in limited transient resolution, as explained in the previous section.
Learning in a direct fashion allows us to explicitly address and reproduce transients during training.

\subsection{Inference}

Model inference takes place within the Tone Transfer pipeline, which takes an input audio signal $i_n$, and yields a synthesized output $s_n$, with $n$ denoting the audio sample index. Similar to other Tone Transfer approaches, we divide this pipeline into three stages:
\begin{enumerate}
\item Feature Extraction, which obtains aligned features from input audio $\hat{a_k}$ and $\hat{f_k}$ related to input amplitude and fundamental frequency $f_{0k}$ respectively, with $k$ denoting frame index. These are extracted from input audio across an analysis window of length $W$. $A(.)$ may denote a signal amplitude or power estimator algorithm, $F(.)$ an $F_0$ tracker, and $G(.)$ a normalization function that maps $F_0$ into the range $[0,1]$.
\begin{equation}
\begin{split}
&\hat{a_k} = A(i_{Wk}, ...i_{W(k+1)}) \\
&f_{0k} = F(i_{Wk}, ...i_{W(k+1)}) \\
&\hat{f_k} = G(f_{0k})
\end{split}
\end{equation}

\item Control Prediction, we use our neural network $g_{\phi}$ to infer a set of frame-wise FM synthesis controls, the oscillator output levels $\hat{ol}_k$, from the conditioning signals $\hat{a_k}$ and $\hat{f_k}$.
\begin{equation}
\hat{ol}_k = g_{\phi}(\hat{a_k},\hat{f_k})
\end{equation}

\item An FM oscillator bank $S_p(.)$ renders a window of $N$ audio samples from output levels $\hat{ol}_k$, fundamental frequency $f_{0k}$. We configure the bank with the oscillator routing and frequency ratios of the patch $p$ used to train $g_{\phi}$, although this can be changed during inference.
\begin{equation}
s_{Nk},...s_{N(k+1)} = S_p(\hat{ol}_k,f_{0k})
\end{equation}
\end{enumerate}

\section{Implementation}

\subsection{Training}
We select a set of common DX7 patches and create a training dataset for each one of them. Next, we train one neural net model per patch following our Envelope Learning method.

For each patch, we generate 1000 random MIDI notes with velocities between 1 and 127, and note values between 0 and 127. We set a random duration for each note between 600 and 732 frames. Next, we generate the aligned input and oscillator envelope sequences $a$, $f$, and $ol$, as described in Section \ref{section:dataset}. Finally, we pad them with zeroes to reach a final size of 1000 frames per instance, so that the active notes occupy about two-thirds of the total length. We split the dataset with a ratio of $[0.80,0.1,0.1]$ for training, validation, and testing respectively. 

Our neural net features a GRU with a hidden size of 128. We empirically choose this value as we note that the training loss does not improve with bigger models, and to keep the computing requirements low. 
We train one model per dataset, for a total of 120000 steps, using the Adam optimizer with a learning rate of 1e-3, a learning rate decay of 0.98 for every 10000 steps, and a batch size of 32 instances. We use Pytorch as a training framework. The process takes about four hours per model using a single NVIDIA GeForce RTX 2080 Ti GPU.

To assess the effectivity of the training process, we compare on the test set the absolute distance between the ground truth oscillator envelopes and the predictions $||ol-\hat{ol}||_1$ for each trained model.

Furthermore, we set out to assess the capabilities of each of the trained networks for synthesizing audio with the learned timbre.
Firstly, we render audio using both the ground truth $ol$ and predicted envelopes $ol$, using an FM oscillator block configured with the oscillator routing and ratios extracted from the patches used to train the models. We employ the fundamental frequency $f_0$ extracted from the normalized MIDI note values present in the sequences $f$ of the test set.

Next, we compute the signal-to-noise ratio (SNR) as a power quotient of our reference and an error computed from the sample-by-sample difference between both signals. We compute the SNR in decibels (dB), as shown in Eqn. \ref{eq:snr}.
\begin{equation}\label{eq:snr}
    SNR = 10 \cdot log_{10} (\frac{||S_p(ol,f_0)^2||_1}{||(S_p(ol,f_0)-S_p(\hat{ol},f_0))^2||_1})
\end{equation}

We use this metric to assess the reconstruction quality of note beginnings and endings.
To account for note beginnings, we aggregate the first 100 milliseconds of each note in the test set for both rendered audios and then compute the SNR on these signals obtaining $SNR_{onset}$. We use 100 ms to account for the different onset times that the models present.
Furthermore, we identify the ending sections of each note by looking at the decaying ramp in $a$, which is aligned with $ol$ in our dataset. We aggregate the audio samples of each note and compute $SNR_{end}$. We aggregate the rest of the audio section of each note, between the note onset and the start of the decaying ramp, and compute $SNR_{mid}$ at the note middle.

Table \ref{table:test_loss} shows the results for the metrics. The low $L_1$ loss indicates that our model is able to minimize the training objective and predict the oscillator envelope sequences from the conditioning signals. This translates into an adequate reproduction of note beginnings, middles and endings; the SNR metrics show that even in the worst case, the models can render the note sections with not more than about 1\% of power error.

Although these results do not represent our models' performance capabilities when deployed in a Tone Transfer pipeline, they show that our training objective allows the networks to learn the envelope contours of the oscillators for different timbres, and can accurately render the beginning, middles, and endings of notes when conditioned with the continuous input sequences $a$ and $f$.

\begin{table}[ht]
\begin{tabular}{|l|l|l|l|l|}
\hline
Model     & Envelope $L_1$ & $SNR_{onset}$ & $SNR_{mid}$ & $SNR_{end}$ \\ \hline
Brass     & 7.77e-4       & 36.4               & 29.3      & 33.7            \\ \hline
Strings   & 1.05e-3       & 29.4               & 34.4      & 34.9            \\ \hline
E. Piano  & 3.06e-3       & 27.5               & 30.2       & 27.9            \\ \hline
Marimba   & 8.07e-4       & 36.5               & 39.7       & 35.8            \\ \hline
Voice     & 7.37e-4       & 19.5                & 33.3     & 30.0           \\ \hline
Sitar     & 2.23e-3       & 22.2                & 25.3      & 27.2            \\ \hline
PolySynth & 1.36e-3       & 30.6               & 37.3    & 30.7           \\ \hline
\end{tabular}\caption{Envelope absolute error and audio SNR at beginnings and endings of the test set notes for all trained models.}\label{table:test_loss}
\end{table}

\subsection{Deployment}

We implement the Tone Transfer pipeline on a real-time audio plugin using JUCE and Libtorch, Pytorch's C++ API. Our prototype can load new neural net models and FM configurations, supporting all the learned timbres. It runs in real-time and performs inference and synthesis at a frame rate of 690 Hz, to render audio at a sample rate of 44.1kHz, similar to our Yamaha DX7 reference implementation \cite{gauthier_Dexed_2023}.

Within the pipeline, we extract frame-wise fundamental frequency $f_{0k}$ using the YIN algorithm \cite{decheveigne_YIN_2002}, using an analysis window of 1024 samples, which yields a minimum detectable frequency of about 90 Hz.
We compute the conditioning signal $f_k$ by converting the fundamental frequency values from Hz to MIDI note value and then applying normalization between $[0,1]$, as shown in Eqn. \ref{eq:fnorm}. Furthermore, when a valid fundamental frequency is not detected, the extractor returns zero.
\begin{equation}\label{eq:fnorm}
    f_k = 12 \frac{log_2(f_{0k}/220)+57.01}{127}
\end{equation}

Next, we supply the continuous amplitude input for our system $a$ from a decibel-scale RMS detector. We employ a compute block over a sliding window $W$, clamping the minimum value to -70dB, and normalizing between 0 to 1.
\begin{equation}\label{eq:rmsnorm}
    a_k = 1 + \frac{1}{70} \cdot
    max(-70,log_{10}(\sum_{W}\frac{i_n^2}{W}))
\end{equation}
Since our datasets (and therefore, our trained models) present a linear amplitude range in $a$, our system tries to match normalized RMS in decibels to envelope variations associated with MIDI velocity.

Next, our model predicts the current envelope values for the FM synth, which are interpolated from frame to sample rate and used for the synthesis process. The fundamental frequency $f_{0k}$ is also linearly interpolated and used to drive the oscillators at the synthesis step.

Furthermore, we reset the model's hidden state to all zeroes when both conditioning sequences are zero. This ensures that the model starts from a known state to process a new incoming note.
The plugin runs on a MacBook Pro 2021 with a USB audio interface running at 44.1 kHz and a hop size of 64 samples, yielding a pipeline delay of 3 ms including buffering.

\section{Discussion}

Our model offers the capability to generate a wide range of timbres on an FM synthesizer by learning the dynamic trajectories of oscillator envelopes reflected in the dataset. Our approach effectively replaces the traditional envelope generator of the DX7 with a recurrent neural network (RNN) that provides continuous controllability instead of MIDI.

In this context, the dataset generation approach serves as the bridge between explicit note beginnings and endings, which are event-based, and the continuous control framework.
When trained with our Envelope Learning method, the network is able to learn and reproduce the rapid beginnings and endings of notes, even without explicit information about note boundaries.

Previous Tone Transfer architectures learn to control synthesizers \textit{indirectly} by minimizing an audio loss of a resynthesis task, using spectrogram losses that act upon long windows of audio. These effectively look at the middle of musical events and present a limited temporal resolution for beginnings and endings.
In contrast, our approach overcomes this limitation by learning a direct correspondence between inputs and synthesis parameters at a control level. This allows for precise rendering of the transient characteristics of the learned timbre, provided we have a representation of that timbre available in the form of a synthesizer patch. We argue these are the first steps for achieving expressive and nuanced synthesizer articulation with Tone Transfer algorithms.

We suggest that these results are encouraging to explore the Envelope Learning technique building further input-output associations.
One possibility would be to align additional input features such as spectral features to other sound characteristics like attack and decay rates.
Another would be to introduce multiple notes per training instance to explicitly model phrasing in context, modeling note events of specific musical instruments in the dataset for a more nuanced control.
Other alternatives include exploring further conditioning choices to assess responsivity in terms of dynamics, modifying the trapezoidal amplitude note model in $a$ to better account for particular instruments and input detectors,
or training using patch data from other synthesizer architectures.

\subsection{Reflections on performances}

As a proof of concept, we record two musicians using our audio plugin in real-time, playing guitar and sax \footnote{The video is available on the \underline{\href{https://fcaspe.github.io/fmtransfer}{supplementary website}}}.
We select these two source instruments since they provide very different volume dynamics and articulations to drive our plugin. We record three models, trained with electric piano, strings, and brass patches respectively.

We informally observe that our Tone Transfer approach can effectively render timbre from the learned patches, including 
note beginnings. This is reflected particularly well in the example of the guitar controlling the electric piano, which shows a bright attack on the beginning of those notes that are not \textit{legato}. 
Next, in the guitar example that plays a string tone, the synthesizer features a slow attack, even though the guitar is plucked, showing that our model does not project input loudness to the output, as DDSP does. 

Note endings are much more difficult to assess since their generation depends on a decaying amplitude envelope presented by the audio input. On the guitar, the decay envelope may be too fast to render the learned note ending before the fundamental frequency cannot be tracked anymore.

For the case of the saxophone as a control source, note beginnings and endings are not that clear, but this is to be expected as it presents a much different amplitude contour, including amplitude modulations that were not accounted for during dataset generation. These may force the model to re-render note characteristics of beginnings or endings, which can be observed when the saxophone controls the electric piano.

On the other hand, we note a lack of dynamic range in the synthesis output. We argue that this is due to the fact that the patches were originally designed to be played with a keyboard with MIDI notes and velocity controls. For the electric piano patch, for instance, we observe that a low-velocity value still produces a signal with high amplitude but less brightness. Redesigning the patches to obtain higher variations in output amplitude and retraining the models may improve the results.

\section{Conclusions}

Transforming the audio of an instrument to a synthetic sound is a challenging task, as it involves a one-to-many relationship. Each instrument has its unique timbral palette, dynamic contour, and articulation possibilities, which can vary significantly even among instruments of the same type. On the other hand, the sound produced by a synthesizer can be highly versatile; and only a subset of the source instrument's characteristics may be desired in the output.

We can argue that there is no definitive "gold standard" that can provide a baseline mapping between an instrument's audio and a synthetic sound: tradeoffs are necessary to find viable solutions.

In this work, we first analyzed current Tone Transfer architectures and identified a tradeoff in their rendering capabilities: these models learn new timbres from audio corpora and can project the input loudness to the output, at the expense of a good resolution of note beginnings and endings which are essential for musical articulation and phrasing.

In light of the analyzed shortcomings, we presented Envelope Learning, a design method where a model learns a set of input-to-synthesis parameters correspondences and accurately replicates note beginnings and endings. The tradeoff, in this case, is on the note middles: we use a simplified musical note model for our dataset generation that does not consider variations in amplitude or pitch during an event. This works well during testing in the training environment but may result in unexpected transitions and reduced dynamic range when used in a Tone Transfer setting, especially with sustaining instruments. We leave for future work an assessment of the performance possibilities of our algorithm and an exploration of techniques to overcome current limitations.

Finally, we implemented a Tone Transfer pipeline in an audio plugin for real-time performance, taking a step towards improving sound diversity and phrasing capabilities for audio-based control of synthesizers. Our system bridges the sonic diversity gap of previous approaches, learning new sounds from a vast number of DX7 patches for which their timbre can now be continuously controlled with musical instruments.
We hope that our work motivates further research in model design with the goal of improving phrasing and articulation in real-time neural synthesizers controlled by musical instruments.

\begin{acks}
This work is supported by the UKRI through the Centre for Doctoral Training in Artificial Intelligence and Music (EP/S022694/1) and a UKRI Frontier Research grant (EP/X023478/1).
\end{acks}

\bibliographystyle{ACM-Reference-Format}
\bibliography{envelope_learning}

\end{document}